\input harvmac
\newcount\figno
\figno=0
\def\fig#1#2#3{
\par\begingroup\parindent=0pt\leftskip=1cm\rightskip=1cm\parindent=0pt
\global\advance\figno by 1
\midinsert
\epsfxsize=#3
\centerline{\epsfbox{#2}}
\vskip 12pt
{\bf Fig. \the\figno:} #1\par
\endinsert\endgroup\par
}
\def\figlabel#1{\xdef#1{\the\figno}}
\def\encadremath#1{\vbox{\hrule\hbox{\vrule\kern8pt\vbox{\kern8pt
\hbox{$\displaystyle #1$}\kern8pt}
\kern8pt\vrule}\hrule}}

\overfullrule=0pt

%
\def\C{{\bf C}}
\def\tilde{\widetilde}
\def\bar{\overline}
\def\Z{{\bf Z}}
\def\T{{\bf T}}
\def\S{{\bf S}}
\def\R{{\bf R}}

\font\zfont = cmss10 

\def\bigone{\hbox{1\kern -.23em {\rm l}}}
\def\ZZ{\hbox{\zfont Z\kern-.4emZ}}

\Title{hep-th/9710065, IASSNS-HEP-97-111}
{\vbox{\centerline{NEW ``GAUGE'' THEORIES }
\bigskip
\centerline{IN SIX DIMENSIONS}
}}
\smallskip
\centerline{Edward Witten\foot{Research supported in part
by NSF  Grant  PHY-9513835.}}
\smallskip
\centerline{\it School of Natural Sciences, Institute for Advanced Study}
\centerline{\it Olden Lane, Princeton, NJ 08540, USA}\bigskip

\medskip

\noindent
More general constructions are given of six-dimensional theories that look at 
low energy like six-dimensional super Yang-Mills theory.  The constructions 
start with either parallel fivebranes in Type IIB, or $M$-theory on
$(\C^2\times\S^1)/\Gamma$ for $\Gamma$ a suitable finite group.
Via these constructions, one can obtain six-dimensional theories
with any simple gauge group, and $SU(r)$ theories with any rational
theta angle.  A matrix construction of these theories is also possible.  
\Date{October, 1997}

\newsec{Introduction}

One of the fascinating results obtained in investigations of string duality
and matrix theory is that there exist in six
dimensions theories that do not have dynamical gravity, but have
many properties of string theory, such as $T$-duality upon toroidal
compactification.  Such theories can be obtained by taking the
weak coupling limit of a system of parallel Type II or heterotic string
fivebranes \ref\seiberg{N. Seiberg,  ``Matrix Description Of $M$-Theory
On $\T^5$ and $\T^5/\Z_2$,'' hep-th/9705221.}.

In some cases, these theories have a quite interesting and exotic
infrared behavior, reducing for example at long distances to the
exotic six-dimensional theory that can be constructed from Type IIB
at an ALE singularity \ref\witten{E. Witten, ``Some Comments On String
Dynamics,'' hep-th/9507121.} 
or via parallel $M$-theory fivebranes \ref\strom{A. Strominger,
``Open $p$-Branes,'' hep-th/9512059.}.  Other 
examples look at long distances like infrared-free super-Yang-Mills
theories.  The focus of the present paper will be on such 
``gauge'' theories.  We will furthermore consider only examples that
have the maximal possible supersymmetry for six-dimensional gauge
theories, namely $(1,1)$ supersymmetry.  (There are also many examples
of such theories with $(0,1)$ supersymmetry, as explored recently in
\ref\bi{J. Blum and K. Intriligator, ``New Phases Of String Theory And 6d
RG Fixed Points  Via Branes At Orbifold Singularities,'' hep-th/9705044.}.)

Sections 2-5 are devoted to a class of examples that can be
constructed from a Type IIB system with parallel NS and D fivebranes,
or alternatively via an $M$-theory construction that we present.
These theories have unitary gauge groups at low energies, but differ
from previously studied  systems with such gauge groups; 
in fact, we compute a specific
difference (involving a spacetime theta angle) in section 5.
Section 6 is devoted to examples with orthogonal and symplectic gauge
groups, which can be constructed from Type IIB or from $M$-theory, and
examples with exceptional gauge groups, including $G_2$ and $F_4$, for
which we have only an $M$-theory construction.  A recent analysis
\ref\vafa{C. Vafa, ``Geometric Origin Of Montonen-Olive Duality,'' 
hep-th/9707131.} 
of $S$-duality in ${\cal N}=4$ super Yang-Mills with non-simply-laced  gauge
groups is closely related to and could be formulated in terms of the
six-dimensional theories studied in section 6.  And finally, in section 7
we reexamine
the $SU(r)$ theories of sections 2-5 from the standpoint of matrix theory,
interpreting them via the Coulomb branch of a certain two-dimensional
$(4,4)$ super Yang-Mills theory. 
Issues related to this disccussion 
have also been discussed recently in \ref\othersethi{S. Sethi, 
``The Matrix Formulation of Type IIB Five-Branes, '' hep-th/9710005.}.

\newsec{$(p,q)$ Fivebrane System}

On the world-volume of a collection of $q$ identical
parallel fivebranes in Type IIB superstring theory,
one sees at low energies 
a six-dimensional super Yang-Mills theory with gauge group $U(q)$.
This is so for either Dirichlet or NS fivebranes.  In the case of
NS fivebranes,
the gauge coupling in this $U(q)$ gauge theory is independent of
the string coupling constant $\lambda.$ 
As a result of this
\seiberg, in the limit that $\lambda$ goes
to zero, although the Type IIB string theory becomes free in bulk, the
theory on the fivebranes does not become free.  To be more precise, it seems
very likely (and is implied by a proposed $T$-duality
\ref\ov{H. Ooguri and C. Vafa, ``Two-Dimensional Black Holes And Singularities
Of CY Manifolds,''
hep-th/9511164.} that will be reviewed and generalized in section 4)
that the center of $U(q)$ decouples as $\lambda\to 0$,
and that in this limit there is a surviving six-dimensional supersymmetric
gauge theory with $SU(q)$ gauge group.  This theory has six-dimensional
$(1,1)$ supersymmetry and is 
a non-chiral theory with 16 supercharges, in common
with all other six-dimensional theories that will be discussed in the present
paper.

It is natural to combine $\lambda$ with the Type IIB theta angle $\theta_B$
to a parameter $\tau =\theta_B/2\pi +i/\lambda$ with values in the upper
half plane.  The Type IIB superstring theory is believed to have
an $SL(2,\Z)$ symmetry that acts by 
\eqn\umbo{\tau\to {a\tau+b\over c\tau+d},}
with $a,b,c,d\in \Z$ and $ad-bc=1$.
The limit $\lambda\to 0$ corresponds to $\tau\to i\infty$.  One might
ask whether the limiting theory, obtained by taking $\tau\to i\infty$, 
depends on $\theta_B$.  We will give some arguments that
it does not.

In this paper, we will study the more general 
case of a collection of $q$ NS fivebranes and $p$ Dirichlet fivebranes.
If one only wishes to determine the low energy gauge group, which is expected
to be independent of $\tau$, one can
apply an $SL(2,\Z)$ transformation, ignore the action of this
transformation on $\tau$, and map the $(p,q)$ fivebrane system
(the two integers will always be respectively the Dirichlet and NS fivebrane
charges) to an $(r,0)$ system, where $r$ is the greatest common divisor of
$p$ and $q$.  The low energy gauge group is hence $U(r)$.   

If, however, we wish to extract a decoupled six-dimensional theory
by taking $\tau\to i\infty$, then we should restrict ourselves
to $SL(2,\Z)$ transformations which commute with this limit.
This means that we should limit ourselves to upper triangular matrices
\eqn\inno{\left(\matrix{ a & b \cr c & d \cr}\right)
=\left(\matrix{ 1 & n \cr 0 & 1 \cr}\right),}
which act by $\tau\to \tau + n$.
Since the action of $SL(2,\Z)$ on the fivebrane charges is
\eqn\himno{\left(\matrix{ p \cr q\cr}\right) \rightarrow
\left(\matrix{ a & b \cr c & d \cr}\right)\left(\matrix{ p \cr q\cr}\right),}
such transformations map $(p,q)\rightarrow (p+nq,q)$.  Hence $p$ can be
uniquely mapped to the region $0\leq p\leq q-1$, and for each $q$, $p$ has
precisely $q$ possible values.

The theory obtained by taking $\tau\to i\infty$ for given $(p,q)$
has low energy gauge
group $U(r)$ ($r$ being again the greatest common divisor of $p$ and $q$).
For $r>1$, this group is non-abelian.  Moreover, the gauge coupling constant
of this theory is independent of $\tau$ for $\tau\to i\infty$ by the
same sort of arguments as for $p=0$.   Hence,  just as in 
\seiberg, the theory obtained in this way cannot be free if $r>1$.
Arguments given below actually imply that it is not free whenever $q>1$.
This is the strongest possible statement of its sort
since for $q=1$, the generalization
to include Dirichlet fivebranes is vacuous, as the transformation
$(p,q)\to (p+nq,q)$ can be used to set $p$ to zero.

It is very plausible (and follows from the duality we will
propose) that the $U(1)$ factor in $U(r)$ actually decouples, and that the
low energy theory is an $SU(r)$ theory.  If so, there is no low energy
gauge group at all in the six-dimensional theory when $r=1$.  The $M$-theory
interpretation to which we turn next nevertheless implies, for $q>1$, that 
non-trivial six-dimensional physics survives as $\lambda\to 0$ even when $r=1$.

\newsec{$M$-Theory Dual}

\nref\bersh{M. Bershadsky, V. Sadov, and C. Vafa, ``$D$-Strings
on $D$-Manifolds,'' Nucl. Phys. {\bf B463},
hep-th/9510225.}
\nref\asen{A. Sen, ``A Note on Enhanced Gauge Symmetries In $M$-
And String Theory,'' hep-th/9707123.}
\nref\hmg{R. Gregory, J. A. Harvey, G. Moore,
``Unwinding Strings And $T$-Duality Of Kaluza-Klein And $H$-Monopoles,''
hep-th/9708086.}
The goal in this section is to propose, and to begin to explore,
an $M$-theory dual of the $(p,q)$ theory.  For $p=0$, such a dual
has been proposed some time ago \ov\ and studied
from various points of view \refs{\bersh - \hmg}.  
One considers Type IIA superstring theory at an $A_{q-1}$ singularity.
Type IIA perturbation theory breaks down in the field of such a singularity,
if all world-sheet theta angles vanish.  A non-perturbative $SU(q)$ gauge
group appears, with a gauge coupling constant that does not vanish
in the limit that the Type IIA coupling constant $\lambda_A$ vanishes.
Upon taking this limit, one gets a surviving six-dimensional theory with
$SU(q)$ gauge symmetry and $(1,1)$ supersymmetry.  This theory is believed
to be equivalent to the one obtained in the weak coupling limit from
$q$ parallel Type IIB fivebranes.  We will review in section 4 a version of the
$T$-duality argument for this equivalence.

We recall that the $A_{q-1}$ singularity can be described as the 
singularity at the origin in the quotient space $\C^2/\Z_q$.  
\foot{To fully exhibit the symmetries, $\C^2$ should here really
be regarded as a flat hyper-Kahler manifold $\R^4$, without a chosen
complex structure.}
Here
$\C^2$ has complex coordinates $z_1,z_2$, and $\Z_q$ acts by
\eqn\obb{\eqalign{ z_1 & \to e^{2\pi i/q}z_1\cr
                   z_2 & \to e^{-2\pi i/q}z_2.\cr}}  

To generalize this dual description of the $(0,q)$ theory to the $(p,q)$ case,
we will have to go to $M$-theory.  Type IIA on $\R^6\times \C^2/\Z_q$
is the same as $M$-theory on $\R^6\times \C^2/\Z_q\times \S^1$.  The weak
coupling limit of Type IIA is the limit that the radius $R$ of the $\S^1 $ is
small; we let $t$, $0\leq t\leq 2\pi R$, be an angular variable on  $\S^1$.

Before going on, let us note a subtlety that will have interesting
analogs later.
Super Yang-Mills theory in six dimensions with $SU(q)$ gauge group
has a moduli space of vacua parametrized by the expectation values of
four scalars $\phi^i$, $i=1,\dots , 4$ in the adjoint representation of
$SU(q)$.  For the vacuum energy to vanish, they must mutually commute
and so can be simultaneously diagonalized.  The $k^{th}$ eigenvalue
$\phi^i_{(k)}$ of the $\phi^i$, for $k=1,\dots,q$, defines a point in $\R^4$.
As the $\phi^i$ are traceless, $\sum_k\phi^i_{(k)}=0$.  
The $\phi^i$ thus define $q-1$ independent
points in $\R^4$.  The moduli space $\cal M$ of
vacua is obtained by dividing by the Weyl group, which is the group $\S_q$
of permutations of the eigenvalues, so one has
\eqn\mommo{{\cal M}=(\R^4)^{q-1}/\S_q.}
Now, $M$-theory on $\C^2/\Z_q\times \S^1$ has a moduli space
which is instead
\eqn\ommo{\tilde{\cal M}=(\R^3\times \S^1)^{q-1}/\S_q.}
It differs from $\cal M$ because one of the four scalars in the
six-dimensional effective super Yang-Mills theory originates from
 the Wilson line
around the $\S^1$ factor in $\C^2/\Z_q\times \S^1$.  This Wilson line
is group-valued, not Lie algebra valued, so its eigenvalues live
in $\S^1$, not in $\R$; taking account of the Wilson line together
with three seven-dimensional scalars in the adjoint representation gives
\ommo.

The circle in \ommo, because it originates as a holonomy,
is the {\it dual} of the original circle in $\C^2/\Z_q\times \S^1$.
So as we take the radius of the original circle to zero, the radius of the
circle that enters in the definition of $\tilde {\cal M}$ goes to
infinity.  Hence, in the decoupled six-dimensional theory, this circle
is decompactified.  To obtain a six-dimensional theory,
one takes the coupling to zero in a vacuum given by some point
$P\in \tilde {\cal M}$, and in the limit of zero coupling, one only sees
a small neighborhood of $P$ (that is, a neighborhood vanishingly
small compared to the
radius of the $\S^1$).  

The limiting theory that one gets depends on the choice
of $P$.  If $P$ is a generic, smooth point in $\tilde {\cal M}$, the
limiting theory is a free abelian theory; if $P$ is the singularity
at which all $q$ points in $\R^3\times U(1)$ coincide, the limiting
theory is an $SU(q)$ theory; if $P$ is a singular point at which
$q'$ points coincide ($1<q'<q$) one gets a theory of the same structure
with $SU(q')$ gauge group.  The appearance here of the $SU(q')$ theory
does not have much novelty because it can be obtained in any case
by first constructing the six-dimensional $SU(q)$ theory and then Higgsing
it.  In that sense then, in the present example everything interesting
comes by considering the case that $P$ is the maximally singular point.
If, however, $\tilde{\cal M}$ had more than one maximally
singular point, then essentially different six-dimensional theories
(no one of which could be obtained by Higgsing another) could arise from
small radius limits of $M$-theory on $\C^2/\Z_q\times \S^1$ for different
$P$.  We will find such  situations later in
this paper. 

\subsec{Generalization}

We propose now to generalize Type IIA at an $A_{q-1}$ singularity
to  $M$-theory on $\R^6\times (\C^2\times \S^1)/\Z_q$, where $\Z_q$ 
acts by
\eqn\nobb{\eqalign{z_1 & \to e^{2\pi i/q}z_1\cr
                   z_2 & \to e^{-2\pi i/q}z_2\cr
                   t & \to t-2\pi R{p\over q}\cr}}
with integer $p$.
We write $X_{p,q}$ for $(\C^2\times \S^1)/\Z_q$ with this action of
$\Z_q$.  Notice that $p$ is only well-defined modulo $q$, and that in 
particular
if $q=1$, $X_{p,q}$ is just the original $\C^2\times \S^1$.
So just as in the previous section, the interesting cases are $q\geq 2$,
$0\leq p < q$.

We want to consider $M$-theory on $\R^6\times X_{p,q}$.
In the limit $R\to 0$,
this becomes if one is not very near $z_1=z_2=0$ equivalent to
weakly coupled Type IIA on $\C^2/\Z_q$.  In fact, an observer at a 
point $P\in \C^2$  that is not close to $z_1=z_2=0$ does not detect by 
local measurements                   
that $P$ has been identified with $q-1$ image points, and so in particular
does not detect the details of how this identification was made.  
So in bulk, $M$-theory
on $\R^6\times X_{p,q}$ becomes a free Type IIA theory for $r\to 0$.

As long as $p$ is not congruent to 0 mod $q$, the theory  near
$z_1=z_2=0$ cannot be a free Type IIA theory, for the following reason.  
As in section 2, let $r$ be the greatest common divisor of $p$ and $q$,
and let $s=q/r$.  Thus, $s>1$ if $p$ is not congruent to 0 mod $q$.
As was noted in \ref\senschwarz{J. Schwarz and A. Sen,
``Type IIA Dual Of The Six-Dimensional CHL Compactification,''
hep-th/9507027.}
in a context closely related to the present discussion, the present
theory has the following unusual property from a Type IIA point of view.
In addition to conventional Type IIA superstrings which can be found at
any values of $z_1,z_2$ and come from $M$-theory membranes that wrap once
around $\S^1$, the theory has additional strings whose tension is
$1/s$ times the standard Type IIA tension and which propagate only
on $\R^6$, that is only at $z_1=z_2=0$.  
The reason for this is that at $z_1=z_2=0$, $\Z_q$ is acting only
on $\S^1$, creating a circle that is $s$ times smaller than the original 
$\S^1$.  So near the origin, there are strings that come from membranes
that wrap once around $\S^1/\Z_s$.  Such strings are prevented from
moving away from a small neighborhood of $z_1=z_2=0$ by a strong 
energetic barrier.  Their tension is $s$ times smaller than the usual
Type IIA string tension because they come from membranes that are $s$ times
shorter. (Like the ordinary strings, the exotic ones are BPS saturated, but
with a charge $s$ times smaller; so the ratio of tensions is precisely $s$.) 
$s$ of these strings  can combine to an object
that can move without energetic cost
to large $z_1,z_2$ and turn into an ordinary Type IIA
superstring near infinity.

The theory of weakly
coupled Type IIA strings cannot be enriched by adding strings of fractional
charge while preserving the existence of a systematic string perturbation
expansion.  So the presence of these fractional strings shows that the theory
with $p$ not congruent to 0 mod $q$ cannot reduce to a free theory
in the limit of small $R$.  It is a free Type IIA
theory in bulk, but there are
surviving interactions at $z_1=z_2=0$.  (Of course, this is also true
for $p=0$, $q>1$ because of the $SU(q)$ gauge symmetry.  This is the more
familiar case that we discussed first.) In this way, we get a family
of six-dimensional theories, with $(1,1)$ supersymmetry, parametrized
by the choice of integers $(p,q)$ with $q\geq 2$ and $p\cong p+q$.

What is the low energy gauge group of this theory?  It is expected to be
independent of $R$ (unless there is a very unfamiliar kind of phase transition)
and is conveniently determined by going to large $R$ and using the 
long-wavelength approximation to $M$-theory.  Non-abelian gauge
symmetry in $M$-theory comes from singularities.  In the present case,
singularities of $X_{p,q}$ come only from fixed points of the $\Z_q$
action on $\C^2\times \S^1$.  Since  $\Z_q$ acts on $\S^1$ by translations,
any element of $\Z_q$ that acts non-trivially on $\S^1$ acts freely on
$\S^1$ and hence on
$\C^2\times \S^1$.  Fixed points and singularities come therefore only
from the subgroup of $\Z_q$ that acts trivially on $\S^1$.  This
subgroup is isomorphic to $\Z_r$.  Its action on $\C^2$ is generated
by 
$z_1\to e^{2\pi i /r}z_1,
\,\,z_2\to e^{-2\pi i/r}z_2$.  
Dividing by this action
produces an $A_{r-1}$ singularity, corresponding in seven dimensions
to $SU(r)$ gauge symmetry.    

Thus along $\R^6\times \S^1$, we have locally an $SU(r)$ gauge symmetry.
All told, from a six-dimensional point of view, we have an $\S^1$ family
of $A_{r-1}$ singularities.  To determine the gauge group in six dimensions,
the remaining question is to determine whether, in going around the circle,
there is a monodromy that breaks $SU(r)$ to a subgroup.  Such a monodromy
would necessarily involve an outer automorphism of $SU(r)$; there is only
one non-trivial outer automorphism, ``complex conjugation.''\foot{Monodromy
consisting of an inner automorphism, that is an element of $SU(r)$, 
means that there is a Wilson line on $\S^1$
breaking $SU(r)$ to a subgroup.  One can always turn on such a Wilson
line before shrinking the radius of $\S^1$, but this does not give
anything essentially new in the following sense.
As in the discussion of
\ommo, this gives the same
six-dimensional theory that one would obtain by first taking the 
small radius limit with unbroken $SU(r)$ and then Higgsing
the six-dimensional theory.}

In general, for a family of $A-D-E$ singularities, there is monodromy,
breaking the  gauge symmetry to a subgroup.  This has been found to play
an important role in $F$-theory \nref\aspinwall{P. Aspinwall and
M. Gross, ``The $SO(32)$ Heterotic String On A K3 Surface,''
Phys. Lett. {\bf B387} (1996) 735, hep-th/9605131.}
\nref\vseveral{ M. Bershadsky, K. Intriligator, 
    S. Kachru, D.R. Morrison, V. Sadov, and C. Vafa,
 ``Geometric Singularities and Enhanced Gauge Symmetries,'' hep-th/9605200.}
 \refs{\aspinwall,\vseveral}, 
and we will meet examples of such monodromy
later.  In the present case, however, the monodromy vanishes and the
gauge group in six dimensions is actually $SU(r)$.  This assertion
can be justified as follows.

$\C^2/\Z_r$, regarded as a flat hyper-Kahler manifold with an orbifold
singularity, actually has a $U(2)$ symmetry group.  Part of the $U(2)$ is
obscured when one picks a particular complex structure (as we have done
in selecting complex coordinates $z_1,z_2$
to identify the flat hyper-Kahler manifold $\R^4$ with $\C^2$),
but part of it is still manifest.  In particular, there is an evident
 $U(1)$ group of symmetries, which we will call $F$:
\eqn\ijb{\eqalign{ z_1& \to e^{i\alpha}z_1\cr
                   z_2& \to e^{-i\alpha}z_2 .\cr }}
Since this group is connected, and $SU(r)$ has no non-trivial outer
automorphisms that can be continuously connected to the identity, $F$
simply commutes with the $SU(r)$ gauge symmetry.
\foot{This argument should really be made in two steps. (1) As there
are no connected outer automorphisms of $SU(r)$, the automorphism
of $SU(r)$ brought about by $F$ is inner.  Its generator is hence the
sum  of a quantity that commutes with $SU(r)$ and an $SU(r)$ generator.
(2) Hence, by defining
correctly the $F$ action on degrees of freedom localized near the origin
(to remove the possible $SU(r)$ generator from the generator of $F$),
$F$ can be defined so as to commute with $SU(r)$.}
Now, the particular family of $A_{r-1}=\C^2/\Z_r$ singularities of interest
here can be described by saying that as one goes around the ${\bf S}^1$,
$\C^2/\Z_r$ is transformed by a $\Z_q$ transformation.
The monodromy of interest is determined by the action of $\Z_q$
on the $SU(r)$ gauge symmetry.
(Only  the quotient $\Z_s=\Z_q/\Z_r$  acts faithfully
on $\C^2/\Z_r$.)  In view of the form of \nobb,
 $\Z_q$ acts on $\C^2$ as a subgroup of $F$,
and hence commutes with $SU(r)$ and, as we claimed,
does not generate a non-trivial outer automorphism of $SU(r)$.
                 
Therefore, we have obtained a family of exotic six-dimensional
theories, with the same $(1,1)$ supersymmetry, the same low energy gauge
group, and parametrized by the same pairs $(p,q)$, as for the 
$(p,q)$ fivebrane system of Type IIB.  This strongly suggests that the
two classes of six-dimensional theories are the same.  In the next
section, we give via $T$-duality further evidence for this interpretation. 

\newsec{$T$-Duality And $(p,q)$ Fivebrane System}

First we recall the analysis via $T$-duality of the relation between
Type IIA at an $A_{q-1}$ singularity and 
the $(0,q)$ fivebrane system.  A useful reference for this discussion is
\hmg.

Applying $T$-duality directly to $\C^2/\Z_q$ is inconvenient.  It requires
picking one of the $U(1)$ symmetries of $\C^2/\Z_q$ and applying $T$-duality
on the orbits. The radius of the
orbits grows  at infinity, so the $T$-dual has
orbits whose size vanishes at infinity; its behavior at infinity 
is difficult to interpret.

However, the six-dimensional theory obtained from weakly coupled
Type IIA at an $A_{q-1}$ singularity is independent of the global
structure of $\C^2/\Z_q$ and depends only on the behavior near the singularity.
It is hence much more convenient to imbed the singularity in a 
four-manifold which has a convenient $\S^1$ symmetry with a circle
of fixed radius at infinity.   Such a solution is the multi
Kaluza-Klein monopole solution:
\eqn\impo{ds^2={U}(dy-A_i\,dx^i)^2 +U^{-1}(d\vec x)^2,}
with 
\eqn\himpo{U=\left(1+\sum_{a=1}^q{S\over 2|\vec x-\vec x_a|}\right)^{-1}.}
Here $\vec x$ is a three-vector, $\vec x_a$ are the monopole positions
for $a=1,\dots, q$, $A=A_idx^i$ is the vector potential due to $q$ Dirac
monopoles at the given points, and $y$ is a periodic variable of period
$2\pi S$.  In the special case that the $\vec x_a$ all coincide,
the solution \impo\ has an $A_{q-1}$ singularity
at $\vec x =\vec x_a$, where also $U=0$. Near infinity it looks like
an $\S^1$ bundle over $\R^3$ (with $y$ and $\vec x$ being coordinates
of $\S^1$ and $\R^3$, respectively), with first Chern class $q$.

This solution has the interpretation of a system of $q$ Kaluza-Klein
monopoles on $\R^3\times \S^1$.  The presence of the monopoles
causes a topological twisting at infinity.  
The $\S^1$ can be seen in the solution \impo\ as the orbits
of the $U(1)$ symmetry that acts by
$y\to y+{\rm constant}$.  We will call this symmetry $F$ (near the origin
it looks just like the symmetry called $F$ in the previous section).

Now make a $T$-duality transformation, on the $F$ orbits, to convert
to a Type IIB description.  The $T$-duality converts Kaluza-Klein
monopoles to NS fivebranes, as one can see by looking at how $T$-duality
acts on the fields that the monopoles and fivebranes create at infinity
\ref\htt{C. Hull and P. Townsend, ``Unity Of Superstring Dualities,''
hep-th/9410167.}.
\foot{This analysis proceeds as follows.
In compactification on a circle, one generates in Type II superstring
theory two $U(1)$ 
gauge fields; one comes from the metric tensor and one from the
two-form field $B$.  $T$-duality exchanges these two gauge fields
(which couple to string momentum and winding, respectively).
Hence it exchanges
the Kaluza-Klein magnetic charge -- which is the magnetic charge
of the $U(1)$ that comes from the metric --
with a flux of $H=dB$.  This flux can, of course, be seen in the explicit
asymptotic formula for the $T$-dual of the Kaluza-Klein monopole.
But the $H$ flux is the NS fivebrane number.  Hence, the
$T$-dual of a Type IIA state with Kaluza-Klein monopole charge $q$ is
a Type IIB state with NS fivebrane charge $q$.}
So the Type IIB dual of the multi-Kaluza-Klein monopole solution is
a configuration of $q$ NS fivebranes on $\R^3\times \S^1$.  The explicit
solution of the low energy field equations describing $q$ NS fivebranes
on $\R^3\times \S^1$ is known in closed form, but will not be needed
here.

{}From $q$ NS fivebranes on $\R^3\times\S^1$, one gets a $SU(q)$ gauge
symmetry when and only when the fivebranes coincide in space.
Type IIA on a multi-Kaluza-Klein monopole spacetime has extended gauge
symmetry when and only when the $\vec x_a$ coincide, at which point there
is an $A_{q-1}$ singularity.  So the enhanced gauge symmetry from an 
$A_{q-1}$ singularity must be mapped to the enhanced gauge symmetry from 
coincident Type IIB fivebranes.\foot{On each side there is an additional
$U(1)$ that is decoupled from the $SU(q)$.  In the Type IIB description
it comes from the center of mass motion of the $q$ fivebranes, and in Type IIA
it comes from a zero mode of the $B$ field that is not supported near
the singularity  .} 

Upon taking the limit as the string coupling constant goes to zero,
it follows that the decoupled six-dimensional theory obtained
from an $A_{q-1}$ singularity in Type IIA is the same as the decoupled
six-dimensional theory obtained from coincident NS fivebranes in Type IIB.

We can now carry out a discussion analogous to that of \ommo.
The configuration space of $q$ NS fivebranes on $\R^3\times \S^1$ is
$(\R^3\times \S^1)^q/\S_q$.  After factoring out the decoupled center of 
mass position, the moduli space for the relative or internal motion is
$(\R^3\times \S^1)^{q-1}/\S_q$.  
However, in the limit that the Type IIB string coupling is taken to zero,
the radius of the $\S^1$  diverges in the relevant units.
\foot{The reason for this is that as the NS fivebrane has a tension
of order $1/\lambda^2$, in order to have energies of order 1 its transverse
fluctuations have a length scale, in string units, of order $\lambda$.
After rescaling units so that the transverse fluctuations are of order
one, the $\S^1$ factor in the moduli space (which in string units
has a size of order 1) is of order $1/\lambda$.  For related issues
concerning fivebranes with transverse circles, see \ref\sethsei{N. Seiberg
and S. Sethi, ``Comments On Neveu-Schwarz Fivebranes,''
hep-th/9708085.}.}  The six-dimensional
theory obtained by taking the coupling to zero while sitting at some point
$P\in (\R^3\times\S^1)^{q-1}/\Z_q$ only ``sees'' a small
neighborhood of $P$.  The most interesting theory, with low energy
gauge group $SU(q)$, is obtained by taking $P$ to be the most singular
point at which all fivebranes coincide in space.  All possibilities
can be obtained by Higgsing this theory.

\bigskip\noindent{\it $(p,q)$ Generalization}

Our goal is to consider the $(p,q)$ case in a similar way.

Let $W$ be the charge one Kaluza-Klein monopole solution:
\eqn\bimpo{ds^2={U}(dy-A_i\,dx^i)^2 +U^{-1}(d\vec x)^2,}
with 
\eqn\bhimpo{U=\left(1+{S\over 2|\vec x|}\right)^{-1}.}
$y$ is a periodic variable of period $2\pi S$; $W$ is a smooth manifold.
Consider the $\Z_q$ action on this space generated by $y\to y+2\pi S/q$.
The quotient $W_q=W/\Z_q$ is a singular hyper-Kahler manifold with
an $A_{q-1}$ orbifold singularity at the origin; it is actually
equivalent to the special case of the multi-Kaluza-Klein monopole
solution \impo\ in which the monopoles are all coincident at the origin
(but with $S$ replaced by $S/q$).

We want to consider the same construction as in section 3, but with
$\C^2$ replaced by $W$ and $\C^2/\Z_q$ replaced by $W_q$.  The change
is inessential as far as the limiting six-dimensional theories
are concerned (since $W$ and $W/\Z_q$ look near the origin like
$\C^2$ and $\C^2/\Z_q$) but makes it possible to use $T$-duality.
So we let $\S^1$ be a circle
with periodic  coordinate $t$ of period $2\pi R$, and
we consider $M$-theory on a quotient
$Y_{p,q}=(W\times \S^1)/\Z_q$, with  $\Z_q$ acting on $W$ as in the last
paragraph and on $\S^1$ by $t \to t  -2\pi R(p/q)$.

We can now make an argument just like that in section 3.  In the limit
of $R\to 0$, $M$-theory on $\R^6\times Y_{p,q}$ looks like free Type IIA
superstring theory, everywhere except near fixed points of the $\Z_q$ action
on $W$ (that is, near $\vec x=0$ in \bimpo).  Near those fixed points,
$Y_{p,q}$ looks like the space $X_{p,q}=(\C^2\times \S^1)/\Z_q$
studied in section 3.  The surviving six-dimensional theory in the limit
$R=0$ should be the same whether one considers $X_{p,q}$ or $Y_{p,q}$.
But (as in the $p=0$ case that was reviewed above) using $Y_{p,q}$ makes
it easier to make a relation to Type IIB.

The relation to Type IIB is made by using the fact that $M$-theory on
$\T^2$ is equivalent to Type IIB on $\S^1$.  Let $N$ be the quotient of
$W$ by the $U(1)$ symmetry $y\to y+{\rm constant}$, which we have called $F$.  
Thus, $N$ is a 
copy of $\R^3$, parametrized by $\vec x$, but actually has
a singularity at $\vec x=0$.  Near infinity, $W$ is
fibered over $N$ with $\S^1$ fibers of circumference $2\pi S$.  
Hence $W\times \S^1$ is
fibered over $N$ with fibers $\T^2=\S^1\times \S^1$; metrically
the $\T^2$ is a rectangular torus that is a product of circles with
circumferences $2\pi S$ and $2\pi R$. One can think of $\T^2$ as
the quotient of the $y-t$ plane by the lattice $\Gamma$ generated by
$e=(2\pi S,0)$ and $f=(0,2\pi R)$. The $\tau$ parameter of this
$\T^2$ is hence
\eqn\jtau{\tau=i{S\over R},}
and as $R\to 0$, $M$-theory on $W\times \S^1$ is equivalent near infinity
to weakly coupled Type IIB on $N\times \S^1$. (The $\S^1$ in this
Type IIB description is
a dual circle with a radius of order $1/RS$.)  Near the origin in $N$,
there is a singularity of some kind in the Type IIB theory.  From our
above discussion of $T$-duality in Type IIA on $W$ (which is the same as
$M$-theory on $W\times \S^1$), this singularity is an NS fivebrane.

Now we want to repeat this for $M$-theory on $Y_{p,q}=(W\times \S^1)/\Z_q$.
This still looks at infinity like a two-torus bundle over $N$.
The fibers are now copies of $\tilde T=\T^2/\Z_q$, where 
$\Z_q$ acts by $y\to y+2\pi S/q$, $t \to t -2\pi R(p/q)$.
$\tilde T$ is the torus obtained by dividing the $y-t$ plane
by the lattice $\tilde \Gamma$
generated by $f=(0,2\pi R)$ and $\tilde e=(2\pi S/q,-2\pi Rp/q)$.
The $\tau$ parameter of this lattice is 
\eqn\roggo{\tau=i{S\over q R}-{p\over q},}
and this is the $\tau$-parameter of the equivalent Type IIB theory.
In particular, for $R\to 0$ we get as before a weakly coupled Type IIB
theory, but now with a theta angle  $\theta_B=-2\pi p/q$.  The significance,
if any, of this theta angle will be the subject of the next section.

One has, evidently,
\eqn\lolly{e=q\tilde e +pf.}
This implies the following.  Consider Type IIB theory
with a $\tau$ parameter associated with a lattice $\R^2/\Gamma$.  The
possible five-brane (or one-brane) charges correspond to lattice
points in $\Gamma$.  Five-brane (or one-brane) tensions are proportional
to the lengths of the corresponding lattice vectors.  For $M$-theory
on $W\times \S^1$ with large $S/R$ interpreted as a weakly coupled
Type IIB theory,  the lattice vectors $e$ and $f$ correspond respectively
to an NS fivebrane and a Dirichlet fivebrane. This spacetime has a charge
corresponding to a single NS fivebrane (as we recalled above via $T$-duality
from Type IIA), which corresponds to the charge vector $e$.  After
dividing by $\Z_q$, the charge vector $e$ is reexpressed in the lattice
appropriate to $M$-theory on $(W\times \S^1)/\Z_q$ via \lolly.
\foot{In general dividing by $\Z_q$ does not commute with quantum dynamics,
but since the charge is determined by the asymptotic ``twisting'' of
the $\T^2$ at infinity, which is a topological invariant,  the
charge vector on $(W\times\S^1)/\Z_q$ is simply that of $W\times\S^1$
expressed in the new lattice.}
It therefore corresponds to a $(p,q)$ fivebrane system of Type IIB.

This strongly supports the claim that the six-dimensional theory extracted
from $M$-theory on $X_{p,q}$ coincides with that obtained from
a $(p,q)$ fivebrane system in weakly coupled Type IIB. 

\newsec{The $\theta$ Angle} 

An interesting feature of the $T$-duality argument made in the last section is
that, in this approach, the Type IIB theory naturally appeared with
a theta angle  of $\theta_B=2\pi {\rm Re}\,\tau=-2\pi p/q$.  Is this
significant? 
This is a special case of the following question: does the physics
of the six-dimensional theories obtained by taking the Type IIB coupling
to zero depend on the Type IIB theta angle?

We will argue, though not conclusively, that the answer is ``no.''
On the other hand, we will claim that these theories {\it do} have
an observable and significant theta angle that depends on $p$ and $q$; it
simply is not the Type IIB theta angle!

In  the discussion so far, for each $r$ ($r\geq 2$),
we have obtained infinitely many 
six-dimensional theories with low energy $SU(r)$ gauge
group.  Indeed, let $(a,b)$ be any pair of relatively prime integers;
upon setting $(p,q)=(ar,br)$ and considering a $(p,q)$ fivebrane system,
we get for each pair $(a,b)$ a theory with low energy gauge group $SU(r)$.
How would a low energy observer distinguish these theories?  We will
consider  the case $r\geq 3$ (though the question is natural
for $r=2$, and even for $r=1$ where there is no low energy gauge symmetry).

The low energy interactions are dominated by the $F^2$ term in the effective
Lagrangian
and its supersymmetric generalization; by considering only these interactions,
one certainly cannot distinguish different theories with the same low
energy gauge group.  It will be necessary to look at interactions of higher
dimension.  

One higher dimension interaction of special significance is the theta
angle of the gauge theory.  For $r>2$, one has $\pi_5(SU(r))=\Z$,
so there is a theta angle in $SU(r)$ gauge theory in six-dimensions.
It could be measured, in principle, by a low energy observer studying
the $(p,q)$ six-dimensional theory.  (Its qualitative consequences are
most readily understood in a topologically non-trivial situation,
either on a general six-manifold or in the presence of instanton
strings on $\R^6$.)  The question arises of what value
of $\theta $ would be observed by a low energy observer probing the
six-dimensional theory extracted from the $(p,q)$ system.

To answer this, we first consider the case of $r$ {\it Dirichlet}
five-branes.  This $(r,0)$ case does {\it not} lead to an interacting
six-dimensional theory in the limit that the Type IIB coupling vanishes,
but of course there is a low-energy $SU(r)$ gauge symmetry for such a system.
Moreover, according to the general theory of couplings of RR fields
to $D$-brane world-volume gauge fields, for the $(r,0)$  system, 
the space-time theta angle is simply equal to the
underlying Type IIB theta angle.  In fact, if $\beta$ is the RR scalar
of Type IIB (so that $\theta_B=2\pi
{\rm Re}\,\tau$ is the expectation value of $\beta$),
then there is a world-volume coupling $\beta\Tr \,F\wedge F\wedge F$,
which upon setting $\beta$ to its vacuum expectation value reduces to
\eqn\ubbu{{\rm Re}\,\tau \,\,\Tr F\wedge F\wedge F.}
The coefficient of this interaction is precisely such
\foot{The RR interactions are written as $\beta \,\,{\rm ch} F$, 
where ${\rm ch}$
is the Chern character\ref\hmgreen{M. Green, J. Harvey, and G. Moore,
``$I$-Brane Inflow And Anomalous Couplings On $D$-Branes,''
Class. Quant. Grav. {\bf 14} (1997) 47,
 hep-th/9605033.}.  The six-form part of the Chern character 
integrates in general for a gauge field of finite action
(or on a compact six-manifold) to an arbitrary integer.  That is why the
normalization of the $\Tr F\wedge F \wedge F$ term
precisely leads to $\theta=2\pi\,\, {\rm Re}\,\tau$ with
no additional numerical factor.}
that the space-time theta angle is 
 $\theta=2\pi {\rm Re}\,\tau$ for this case.  

We can map the $(r,0)$ system to a $(p,q)=(ar,br)$ system via an
$SL(2,\Z)$ transformation by a matrix $\left(\matrix{ a & e\cr b & f}\right)$
for some $e,f$.  (Suitable $e$ and $f$ exist  because $a$ and $b$ 
are relatively prime.)  This maps 
\eqn\hotto{\tau\to \tau'={a\tau +e\over b\tau+f}.}
To get a decoupled six-dimensional theory, we need ${\rm Im}\,\tau'\to\infty$,
and this evidently happens for $\tau\to -f/b$.  Since
the $\theta$ angle is $2\pi {\rm Re}\,\tau$, we get
$\theta=-2\pi (f/b)$.  (The sign actually depends on conventions
that we will not try to fix precisely.)

Thus a low energy observer can distinguish the different theories with the
same gauge group by measuring $\theta$.  Now the question arises
of whether only rational values of $\theta$ are possible in the low
energy theory, or whether a generalization exists with irrational $\theta$.
A proposal for a generalization of the construction discussed
in the present paper to get variable  $\theta$ has been made 
recently by Kol \ref\kol{B. Kol,
``On 6-D `Gauge' Theories With Irrational Theta Angle,'' hep-th/9711017.}.

Note that the $\theta$ angle of the low energy theory depends only on the
fivebrane charges, and not on an additional parameter such as ${\rm Re}\,
\tau'$.  This gives some evidence that ${\rm Re}\,\tau'$ is not
an observable parameter in the theory obtained by taking 
${\rm Im}\,\tau'\to\infty$.

\newsec{Other Gauge Groups}

We would like to generalize this discussion to consider six-dimensional
theories with gauge groups other than $SU(n)$.  We begin by analyzing
theories with orthogonal and symplectic gauge groups, followed by
a brief discussion of exceptional groups.

\subsec{Orientifolds And Their Cousins}

First we recall how orthogonal and symplectic
gauge groups are obtained in Type IIB by using Dirichlet fivebranes.
The basic idea is to consider such fivebranes at an orientifold sixplane.
Thus, we consider Type IIB on $\R^4/\rho\Omega$, where 
 $\rho$ is a reflection of all four coordinates,
and the action of $\rho$ is combined with exchange of world-sheet left
and right-movers, which we call $\Omega$.  Away from the origin,
$\R^4/\rho\Omega$ looks just like $\R^4/\rho$, but the behavior near the
origin is different.

Since $\Omega$ reverses the sign of $\theta_B$,
orientifolding of Type IIB can conceivably be carried out either at
$\theta_B=0$ or at $\theta_B=\pi$.  However, only the constructions
at $\theta_B=0$ are well-understood.
At $\theta_B=0$, there are two kinds of orientifold sixplane for Type IIB.
They differ by the sign of the ${\bf RP}^2$ contribution in string
perturbation theory.  We will call the two types of sixplane
${\cal O}^+$ and ${\cal O}^-$.  
If one brings $n$ Dirichlet fivebranes to the fixed
point at the origin in $\R^4$,\foot{I count the number of fivebranes
as measured on the covering space $\R^4$; on $\R^4/\rho\Omega$ one sees half
as many.} then the action of $\Omega$ on the Chan-Paton
factors is symmetric or anti-symmetric in the two cases, leading
to orthogonal gauge groups for ${\cal O}^+$ and symplectic ones for
${\cal O}^-$.  For ${\cal O}^+$, the number of fivebranes can be either
even or odd,
giving an $SO(n)$ gauge group with even or odd $n$.  
Fivebranes can only move to or from the origin in pairs, so the number
of fivebranes at the origin is conserved modulo two.
For ${\cal O}^-$, the
number of fivebranes must be even, $n=2m$, and the gauge group is $Sp(m)$.

The orientifold six-planes carry Dirichlet fivebrane charge $-2$ in the
case of ${\cal O}^+$, and $+2$ in the case of ${\cal O}^-$.  The charge
is reversed in going from ${\cal O}^+$ to ${\cal O}^-$ because the
sign of the ${\bf RP}^2$ contribution to the world-sheet path integral
is reversed.\foot{The numerical value is obtained as follows.
In general,  an orientifold $10-k$-plane has $9-k$-brane charge
$\mp 32(2^{-k})$; this is $\mp 2$ for six-planes.}

{}From Dirichlet fivebranes one cannot make a decoupled six-dimensional
theory.  To do so, we should consider NS fivebranes.  So we make
an $S$-duality transformation $\tau\to -1/\tau$, which converts
the D fivebranes to NS fivebranes, and also maps $\Omega$ to $(-1)^{F_L}$
-- the operation that counts left-moving world-sheet fermions modulo two.
The space-time is now $\R^4/\rho (-1)^{F_L}$; actually, there are
necessarily two versions ${\cal U}^\pm$
of $\R^4/\rho\cdot(-1)^{F_L}$, obtained by $S$-duality
from ${\cal O}^+$ and ${\cal O}^-$, respectively.  They have NS fivebrane
charges of $-2$ and $+2$, respectively.
When $n$ NS fivebranes approach the origin in $\R^4/\rho\cdot(-1)^{F_L}$,
one gets a gauge group $SO(n)$ or $Sp(n/2)$ for ${\cal U}^+$ or ${\cal U}^-$.
In the limit that the Type IIB string coupling constant $\lambda$
vanishes, one
gets six-dimensional theories with these gauge groups, decoupled from the
bulk.

\def\U{{\cal U}}
In comparing to $M$-theory, it will be more convenient to 
replace $\R^4/\rho(-1)^{F_L}$ by $(\R^3\times \S^1)/\rho(-1)^{F_L}$
(where now $\rho$ acts as $-1$ on all three coordinates of $\R^3$ and also
on an angular coordinate on $\S^1$).  The reason, as in section 4, is
that this facilitates arguments using standard dualities.
Since there are two fixed points in the $\rho$ action on $\R^3\times \S^1$,
there are several distinct cases; the fixed points may be
$\U^+\U^+$, $\U^+\U^-$, or $\U^-\U^-$.  
To obtain a decoupled six-dimensional theory, one takes $\lambda\to 0$, 
keeping the radius of the $\S^1$ fixed in string units.  As in a footnote
in section 4, the scale
of variation of the fivebrane positions is of order $\lambda$, so for
$\lambda\to 0$, one does not ``see'' the whole $\S^1$, but only
a vanishingly small neighborhood of the vacuum, which is determined
by the fivebrane positions.  Hence, in general 
inequivalent six-dimensional theories  can be reached by taking
$\lambda\to 0$ with  different fivebrane positions; these six-dimensional
theories will be essentially different in the sense that no one of them
can be obtained by Higgsing another.
The main examples are as follows.
By $n$ fivebranes at a generic point, one gets for $\lambda\to 0$
the $U(n)$ theory
studied above, which we will call $P_n$; 
from $n$ fivebranes at a ${\cal U}^+$ fixed point,
one gets an $SO(n)$ theory that we will call $Q_n$; 
from $n=2m$ fivebranes at a  ${\cal U}^-$ point, we get  an $Sp(m)$ theory
that we will call $R_m$.  Other examples are simple consequences of these.

Here are some cases whose $M$-theory counterparts we will find later.
$n$ will denote the total number of fivebranes.
We recall that the number of fivebranes at a ${\cal U}^+$ fixed point
can be even or odd, and is conserved modulo two.  In the following
examples, there are several maximally singular configurations, so in
the weak coupling limit, one can obtain distinct six-dimensional theories
that cannot be Higgsed to one another.

(1) In the $\U^+\U^+$ case, if $n$ is even and  the number  of fivebranes
at  each $\U^+$ is even, then by placing $2s$ fivebranes at one fixed point and
$n-2s$ at the other, we make for $\lambda\to 0$ the product theory 
$Q_{2s}\times Q_{n-2s}$ with low energy gauge group $SO(2s)\times SO(n-2s)$.  
No one of these theories can be Higgsed to another.
Note that the total NS fivebrane charge of the orientifolds is $-4$, so the
total fivebrane charge is $n-4$.

(2) In the $ \U^-\U^+$ case, by placing $2s$ fivebranes at $\U^-$ and
$n-2s$ at $\U^+$, we make the theory $R_s\times Q_{n-2s}$ with gauge group
$Sp(s)\times SO(n-2s)$; these theories
cannot be Higgsed to one another.  The total NS fivebrane charge of the
orientifolds is 0, so the total fivebrane charge is $n$.  This example
makes sense whether $n$ is even or odd.

(3) In the $\U^+U^+$ case, if $n$ is even and the
number of fivebranes at each $\U^+$ is odd,
then by placing $2s+1$ at one fixed point and $n-2s-1$ at the other,
we make the theory $Q_{2s+1}\times Q_{n-2s-1}$ with gauge group $SO(2s+1)\times
SO(n-2s-1)$; these theories cannot
be Higgsed to one another.  The total NS fivebrane charge of the orientifolds
is $-4$, so the total fivebrane charge is $n-4$.

\subsec{Approach Via $M$-Theory}

Now we will approach the same subject via $M$-theory.

To start with, we recall how to get in seven dimensions a gauge
group $D_k=SO(2k)$.  For this, one considers $M$-theory on $\R^7\times \C^2/
\Gamma$, where $\Gamma$ is generated by
group elements $\alpha, \,\beta$, obeying 
\eqn\noggo{\eqalign{ \alpha^2 &=\beta^{k-2} \cr
                      \alpha\beta & =\beta^{-1}\alpha\cr
                        \alpha^4&=\beta^{2k-4}=1.\cr}}                        
It follows from the first relation that $\alpha^2$ commutes with both
$\alpha$ and $\beta$ and hence is a central element of the group, whose
square is the identity according to the last relation.  (The quotient of
$\Gamma$ by the group generated by $\alpha^2$ is a dihedral group.)

The action of $\Gamma$ on $\C^2$ is
\eqn\poggo{\eqalign{\alpha:& \left(\matrix{z_1 \cr z_2\cr}\right)
    \to \left(\matrix{z_2 \cr -z_1\cr}\right) \cr
    \beta:& \left(\matrix{z_1 \cr z_2\cr}\right)
    \to \left(\matrix{e^{\pi i/(k-2)}z_1 \cr e^{-\pi i/(k-2)}z_2}\right) .\cr}}

We can obtain an $SO(2k)$ gauge theory in six dimensions
by considering $M$-theory on $\R^6\times \C^2/\Gamma\times \S^1$,
where $\S^1$ has circumference $2\pi R$.  This theory is in fact equivalent
to Type IIA on $\R^6\times \C^2/\Gamma$.  The Type IIA string coupling
constant vanishes for $R\to 0$, so the theory becomes free in bulk, but
the $SO(2k)$ gauge coupling, if expressed in string units, is non-zero
in this limit.\foot{For example, gauge  fields that gauge the maximal
torus of $SO(2k)$ are RR gauge fields of the perturbative Type IIA
theory, and their gauge couplings are independent of the string coupling
constant \ref\oldwitten{E. Witten, ``String Theory Dynamics In Various
Dimensions,'' Nucl.Phys. {\bf B443} (1995) 85, hep-th/9503124.}.} 
  So the limiting theory is a six-dimensional theory
with a low energy gauge group $SO(2n)$.                  
                      
As in section 3, we will generalize $\C^2/\Gamma\times \S^1$
to $(\C^2\times \S^1)/\Gamma$, where $\Gamma$ acts as in 
\poggo\ on $\C^2$ while also acting on $\S^1$.  Because in the present
paper, we wish to consider only theories with $(1,1)$ supersymmetry in 
six dimensions, we restrict ourselves to the case that $\Gamma$ acts
on $\S^1$ by rotations (inclusion of reflections would break half the
supersymmetry).  Since the group of rotations of $\S^1$ is abelian,
$\alpha$ and $\beta$ commute as rotations, and
hence the relation $\alpha\beta=\beta^{-1}\alpha$ reduces to $\beta^2=1$.

There hence are two types of examples:

(A) $\beta$ can act trivially on $\S^1$, in which case $\alpha^2=1$,
so $\alpha$ is a rotation by either zero or $\pi$.

(B) $\beta$ may act by rotation by $\pi$, in which case the relation
$\alpha^2=\beta^{k-2}$ means that $\alpha$ is a rotation by $0$ or $\pi$ if
$k$ is even, or by $\pm \pi/2$ if $k$ is odd.  Whether $k$ is even or odd,
the two choices of $\alpha$ are equivalent, because $\alpha$ can be conjugated
to $\alpha\beta$ by a transformation of the form $z_1\to e^{i\pi/2(k-2)}z_1,
\, z_2\to e^{-i\pi/2(k-2)}z_2$.

Either type of example can be given a Type IIB description by reasoning
along
the lines of section 4.  One first replaces $\C^2$ by the Kaluza-Klein
monopole space $W$.  $W$ has $SU(2)$ symmetry (which acts by rotation
of $\vec x$), and the embedding of $\Gamma$ in $SU(2)$ enables
us to view $\Gamma$ as a symmetry group of $W$.  As such, $\Gamma$ acts freely
except for a fixed point at $\vec x=0$, near which the action of $\Gamma$
on $W$ looks like its action on $\C^2$.  Thus, to study possible
six-dimensional theories supported near the fixed point, one may
replace $\C^2$ by $W$.

Having done so, we proceed as in section 4.  To construct 
$(W\times\S^1)/\Gamma$,
we first divide $W\times \S^1$ by the group $\Gamma'$ generated by $\beta$.
This group is isomorphic to $\Z_{2k-4}$.  As seen in section 
4, $M$-theory on $(W\times \S^1)/\Gamma'$ has an interpretation in terms
of Type IIB compactification on $N\times \S^1$  ($N=W/U(1)$ is a copy
of $\R^3$, parametrized by $\vec x$, but with a singularity at the origin)
with a theta angle that is $\theta_B=0$ in case (A) ($\beta$ acts
trivially on $\S^1$) or $\theta_B=\pi$ in case (B) ($\beta$ generates
a $\pi$ rotation of $\S^1$). 
If we divide only by $\Gamma'$, we get a model that has a Type IIB description,
explained in section 4, with $2k-4$ NS fivebranes and either 0 or $k-2$
$D$ fivebranes, for case (A) or case (B).

What about the effect of dividing by $\alpha$?  A key point is that
$\alpha$ anticommutes with the $U(1)$ symmetry, which we have called
$F$, that generates
the fibers of $W\to N$.  Hence on the two-torus fibers $\T^2=\S^1\times \S^1$
of $W\times \S^1$ (as in section 4, the first factor in $\T^2$
is the orbit of $F$
 and the second factor in $\T^2$ is the second factor in $W\times\S^1$),
$\alpha$ reverses the orientation of the first factor and preserves the
orientation of the second factor.  $\alpha$ thus acts as a $2\times 2$ matrix
\eqn\kiggu{\left(\matrix{ -1 & 0 \cr 0 & 1\cr}\right),   }
generating an outer automorphism of the $SL(2,\Z)$ symmetry group of
Type IIB string theory.  In perturbative Type IIB string theory, this
particular symmetry is seen as $(-1)^{F_L}$, the transformation that counts
left-moving world-sheet fermions modulo two.\foot{The fact that the first
eigenvalue of $\alpha$ is $-1$ and the second is $+1$ means that
$\alpha$ reverses the sign of Dirichlet fivebrane charge and commutes
with NS fivebrane charge, and so can be identified as $(-1)^{F_L}$.}
Note that this transformation changes the sign of
$\theta_B$ (while leaving fixed the string coupling constant), so it is
a symmetry only in the two cases $\theta_B=0$ and $\theta_B=\pi$,
the two values that we are actually encountering.

What about the action of $\alpha$ on the spacetime $N\times \S^1$ of the
Type IIB description?  $\alpha$ reverses the orientation of the $\T^2$
in the $M$-theory description, so (as $M$-theory
membrane wrapping on $\T^2$ becomes
momentum on $\S^1$ in Type IIB) it reverses the orientation of the
$\S^1$.  Since Type IIB does not admit orientation-reversing symmetries,
$\alpha$ must also reverse the orientation of $N$.  Since $\alpha$ must
also commute with rotations of $\vec x$, it must act by $\vec x\to -\vec x$.
(Of course, these assertions
 can also be verified directly, using the form of the Kaluza-Klein
monopole solution).  

Combining these results, $\alpha$ acts on the (A) models
as a reversal of all four
coordinates of $\R^3\times \S^1$, together with $(-1)^{F_L}$.  Thus
$\alpha$ is the    transformation $\rho(-1)^{F_L}$ that we encountered
in our Type IIB discussion above.

\bigskip\noindent{\it Gauge Groups Of ${\rm (A)}$ Models}

Let us now determine, from an $M$-theory point of view,
the low energy gauge groups of the (A)  models.
These are constructed from
 $M$-theory on $\R^6\times (W\times \S^1)/\Gamma$ (or simply
 $\R^6\times (\C^2\times \S^1)/\Gamma$)
with $\alpha$ rotating the $\S^1$ by zero or $\pi$.
Thus we consider two cases.

 If the rotation angle is zero, this theory   is simply
$M$-theory on $\R^6\times W/\Gamma\times \S^1$.
This is equivalent to Type IIA on
$\R^6\times W/\Gamma$, and has gauge group $SO(2n)$ if the Wilson
line around $\S^1$ is trivial.  More generally, the Wilson line around
$\S^1$ can break $SO(2n)$ to a subgroup.  Interesting special cases
are that the Wilson line is a diagonal matrix with $2s$ eigenvalues
$-1$ and $2n-2s$ eigenvalues $+1$.  (The number of $-1$'s must be even,
as the Wilson line is an element of $SO(2n)$.)  By picking this
configuration and taking the limit $\lambda\to 0$, one obtains
a six-dimensional theory with gauge group $SO(2s)\times SO(2n-2s)$.
Once one has flowed to a six-dimensional theory decoupled from the bulk,
these theories cannot be Higgsed to one another (as the scalar fields
are Lie algebra valued and not group-valued, or more rigorously
 the moduli are
$\R$-valued and not $\S^1$-valued).  

This theory should be dual to Type IIB on $(\R^3\times \S^1)/\rho(-1)^{F_L}$,
with NS fivebrane charge $2k-4$.  ($2k-4$ is the fivebrane charge produced
by dividing by the group $\Gamma'$ generated by $\beta$; this group
is isomorphic to $\Z_{2k-4}$.)  Making an $S$-duality transformation
to $D$-fivebranes on $(\R^3\times \S^1)/\rho\Omega$,
we conclude that the results found in the last
paragraph should match one of the examples given at the end of section
6.1 with  Dirichlet
fivebrane charge $2k-4$.  (Our analysis of the duality between $M$-theory
and Type IIB
involved mainly the behavior at infinity and was not precise enough
to predict which kind of orientifold planes should be placed at the
fixed points in the Type IIB description, so we must search by hand among
the examples given in section 6.1.)
Inspection shows that the $M$-theory results agree with example (1)
at the end of section 6.1 if $n=2k$.
So we propose that these
examples match, and as the most interesting consequence that
the six-dimensional $SO(2k)$ theories obtained
 from $M$-theory on $\C^2/\Gamma\times
\S^1$ are the same as those obtained from Type IIB with $2k$ NS
fivebranes at a ${\cal U}^+$ orientifold sixplane.  

This essentially familiar result has been described in some detail to 
facilitate the analysis of subsequent examples.  Let us now consider
the second model of type (A), in which $\alpha$ acts by a $\pi$ rotation
on $\S^1$.

Let as before $\Gamma'$ be the subgroup of $\Gamma$ generated by $\beta$.
Since $\alpha$ acts freely on $W\times \S^1$, singularities are
generated only by the action of $\Gamma'$.  The origin in $W$ is
an isolated fixed point for the action of $\Gamma'$.  As $\Gamma'$ is
isomorphic to $\Z_{2k-4}$, $W/\Gamma'$ has an $A_{2k-5}$ singularity
at the origin.  Since $\Gamma'$ acts trivially on $\S^1$,
$(W\times \S^1)/\Gamma'=W/\Gamma'\times \S^1$ has a
circle of $A_{2k-5}$ singularities.  Locally along $\S^1$, there is
therefore an $SU(2k-4)$ gauge group.  The fivebrane charge, produced
by dividing $W$ by $\beta$ and then reinterpreting in Type IIB language, 
is $2k-4$ as in the previous example.

However, once we divide also by $\alpha$, which rotates $\S^1$ and also
acts by an automorphism of $W/\Gamma'$, we get a {\it non-trivial}
family of $A_{2k-1}$ singularities fibered over $\S^1$.  The question arises
of whether the monodromy of this family acts by an outer automorphism
of $SU(2k-4)$.
In fact, precisely this sort of monodromy of a family of $SU(2n)$ singularities
is encountered in $F$-theory, where it is important that this monodromy
is indeed a non-trivial outer automorphism of $SU(2k-4)$
\refs{\aspinwall,\vseveral}.  In fact, modulo
inner automorphisms (which correspond to turning on Wilson lines), it is
the unique outer automorphism of $SU(2k-4)$; it acts by charge conjugation,
and will be called $\Psi$.  

When an $SU(2k-4)$ group element $g$ is transported around the $\S^1$,
it returns to $\Psi(g)=\bar g$ if there is no Wilson line on the $\S^1$.
If on the other hand there is a Wilson line corresponding to an element
$a\in SU(2k-4)$,\foot{An interplay between Wilson lines and outer
automorphisms similar to what follows has recently been uncovered in $F$-theory
\ref\akm{P. S. Aspinwall, S. Katz, and D. R. Morrison, to appear.}.}
 then the monodromy is $g\to a\bar g a^{-1}$.
The unbroken gauge group consists of group elements such that
\eqn\udd{g=a\bar g a^{-1}.}
This unbroken gauge group depends on $a$.  For $a=1$, the unbroken gauge
group is $SO(2k-4)$.  Suppose on the other hand that
$a$ is block diagonal, with $s$ blocks that are copies of 
\eqn\omiygo{\left(\matrix{0 & 1\cr -1 & 0 \cr}\right)}
as well as a $2k-4-2s$-dimensional identity matrix.
Then the unbroken gauge group is $Sp(s)\times SO(2k-4-2s)$.
These gauge groups, and the total fivebrane charge of $2k-4$, agree
with example (2) of section 6.1 if we set $n=2k-4$.

The most interesting case is to set $s=k-2$.  We learn that
from $M$-theory on $(\C^2\times \S^1)/\Gamma$, in the limit that
the radius of the $\S^1$ goes to zero, we can extract a six-dimensional
theory with gauge group $Sp(k-2)$, and that this is the same theory
constructed in Type IIB from $2k-4$ NS fivebranes at 
a singularity $\R^4/\rho(-1)^{F_L}$ of type ${\cal U}^-$.

\bigskip\noindent{\it Mathematical Formulation}

Before going on, 
it will be useful to put the above 
discussion in a standard mathematical framework,
as we will encounter several additional examples with the same basic
structure.  By mapping $\beta$ to the trivial rotation of $\S^1$ and
$\alpha$ to a rotation by $\pi$, we have defined a homomorphism from 
$\Gamma$ to a subgroup $\Z_2$ of the group of rotations of $\S^1$.
The kernel of this homomorphism is $\Gamma'$, the group generated by $\beta$.  
Thus there is an exact
sequence
\eqn\hff{0\to \Gamma'\to \Gamma\to \Z_2\to 0.}
In the $\Gamma$ action on $\C^2\times \S^1$, the subgroup
$\Gamma'$ acts only on $\C^2$, with a fixed point at the
origin, and so generates a circle of fixed points in $(\C^2\times 
\S^1)/\Gamma$.
The quotient $\Z_2$ acts freely on $\C^2\times \S^1$ (since it acts
freely on $\S^1$).  So the local structure is a singularity of type $\Gamma'$.
Globally, one has a circle of $\Gamma'$ singularities, with a monodromy
generated by the quotient $\Z_2$.

In generalizations, $\Gamma$ will be replaced by a possibly different
finite subgroup of $SU(2)$, $\Gamma'$ will be a subgroup of $\Gamma$,
$\Z_2$ will be replaced by a group $\Z_n$ of
rotations of $\S^1$, for some $n$, and \hff\ will be replaced by
an exact sequence
\eqn\nff{0\to \Gamma'\to \Gamma\to \Z_n\to 0.}
We will consider $M$-theory on $(\C^2\times \S^1)/\Z_n$, where $\Gamma'$
acts only on $\C^2$, but $\Gamma$ acts also on $\S^1$ via the homomorphism
to $\Z_n$.  The singularity in $(\C^2\times \S^1)/\Z_n$ is a circle
of $\Gamma'$ singularities, with $\Z_n$ monodromies, just as in the above
examples.  The possibilities for $\Gamma$, $\Gamma'$, and $n$ have
in fact been classified by Reid in \ref\reid{M. Reid, ``Young Person's Guide To
Canonical Singularities,'' in {\it Algebraic Geometry,} Proc. Symp. Pure
Math. {\bf 46} (1985) 345.}, p. 376.
There are six examples, of which we have so far analyzed two.
Reid's example (1) corresponds to the $(p,q)$ fivebrane system studied in
sections 2-5, while his example (3) (with $n+2$ identified with what
we have called $k$) is the one we have just examined.

\bigskip\noindent{\it The {\rm (B)} Models}

As explained above, there is up to conjugation only one (B) model for
each $k$, but the details depend somewhat on whether $k$ is even or odd.
First we assume $k=2p$ even.  This is Reid's example (5).

In this case, we can suppose that $\alpha$ acts trivially on $\S^1$,
while $\beta$ rotates $\S^1$ by $\pi$.  The subgroup $\Gamma'$ is hence
generated by $\alpha$ and $\gamma=\beta^2$, with relations
\eqn\hocco{\eqalign{\alpha^2& =\gamma^{p-1}\cr
             \alpha\gamma & =\gamma^{-1}\alpha\cr
                  \alpha^4& =\gamma^{2p-2} .\cr}}
$\Gamma'$ is the finite subgroup of $SU(2)$ related to $D_{p+1}=SO(k+2)$.
The action of $\Gamma$ on $\C^2\times \S^1$ is determined by an exact sequence
\eqn\blocco{0\to \Gamma'\to \Gamma\to \Z_2\to 0.}
The global structure is thus a circle of $\C^2/\Gamma'$ singularities,
with $\Z_2$ monodromy.  In a Type IIB description, the fivebrane charge,
being produced by dividing by $\gamma$ (which generates a cyclic
group of order $2p-2$) is $2p-2=k-2$.

The $\Z_2$ monodromy is modulo inner
automorphisms the unique outer automorphism of $SO(k+2)$.
One can take it to be generated by
a reflection on one of the $k+2$ coordinates.  With a suitable
choice of Wilson line, this breaks $SO(k+2)$ to $SO(2r+1)\times SO(k+1-2r)$
(and various subgroups to which one of these can be Higgsed). 
These gauge groups, and the fivebrane charge, agree with example (3)
at the end of section 6.1 if we set $n=k+2$.

We learn from this that a six-dimensional theory with gauge group $SO(2r+1)$
can be obtained from a small radius limit of $M$-theory on
$(W\times \S^1)/\Gamma$, with suitable Wilson lines, and that this theory
coincides with what is obtained from Type IIB with $2r+1$ NS fivebranes
at a singularity $\R^4/\rho(-1)^{F_L}$ of type ${\cal U}^+$.

Finally, we consider the model of type (B) with $k$ odd, say $k=2p+1$.
This correspond to Reid's example (2).\foot{A misprint was pointed out
by D. Morrison; $D_{2n+1}$ should read $D_{2n+3}$.}
In this case, $\alpha$ is a $\pi/2$ rotation.  The group $\Gamma'$ that acts
trivially on $\S^1$ is generated by $\gamma=\beta^2$, with $\gamma^{2p-1}=1$.
The exact sequence is
\eqn\urpo{0\to \Gamma'\to \Gamma\to \Z_4\to 0,}
where one can regard $\Z_4$ as the group generated 
by $\alpha$.  

Since $\Gamma'$ is isomorphic to $\Z_{2p-1}$, the fivebrane charge in
a Type IIB description is $2p-1$.  In $M$-theory on $(\C^2\times \S^1)/\Gamma$,
the global structure
is a circle of $A_{2p-2}$ singularities with monodromies generated by $\Z_4$.
Actually, $A_{2p-2}$ is $SU(2p-1)$, and the monodromy is such that the 
generator
$\alpha$ of $\Z_4$ acts by complex conjugation, while $\alpha^2$ induces
a trivial monodromy.  ($\Z_4$ could not act faithfully, because the group
of outer automorphisms of $SU(2p-1)$ is $\Z_2$.)

As in \udd, the unbroken gauge group depends on a Wilson line $a$.
If $a$ is the sum of $s$ copies of the $2\times 2$ matrix in \omiygo\
plus an identity matrix, then the unbroken gauge group is $Sp(s)\times
SO(2p-1-2s)$.  These possibilities agree with example (2) in section 6.1, with
$n$ now being the odd number $n=2p-1$.  We conclude that $M$-theory
on $(\C^2\times \S^1)/\Gamma$, with this $\Gamma$ action, can give,
in a small radius limit with suitable Wilson lines, a variety of 
$SO$, $Sp$, and $SU$ theories.

We thus obtain, in particular, a second construction of $Sp(n)$ theories
in six dimensions.  Since $\pi_5(Sp(n))=\Z_2$, such theories can have
a $\Z_2$-valued spacetime
theta angle, leading one to wonder if the two $Sp(n)$
constructions give theories with different values of $\theta$, as we argued
in section 5 for $SU(n)$ theories.  Here is a heuristic
argument that the $Sp(n)$ theory made from the (B) model has $\theta=\pi$
while the other one has $\theta=0$.  The (B) model has $\theta_B=\pi$
and fivebrane charges $(p,q)=(k-2,2k-4),$ as we saw above.
To determine the spacetime theta angle, we proceed as in section 5.
We make an $SL(2,\Z)$ transformation
\eqn\nobbo{\tau\to\tilde \tau={\tau\over -2\tau+1}}
which maps the charge vector $(k-2,2k-4)$ to $(k-2,0)$, giving us
therefore a collection of $k-2$ Dirichlet fivebranes.  Reasoning
as in section 5, the spacetime theta angle is hence $\theta=2\pi
\lim_{\tau\to i\infty}{\rm Re}\,\tilde \tau=\pi$.
On the other
hand,  $Sp(n)$ models derived from models of type (A) would by
the same reasoning have $\theta=0$, as in this case the fivebrane charge
vector is $(0,2k-4)$ and the modular transformation is $\tau\to -1/\tau$.

\subsec{Exceptional Groups}

It remains to briefly discuss the cases that involve exceptional
gauge groups.

There are two more cases of exact sequences along the 
lines of \nff.  (They are Reid's examples (4) and (6).)
One reads
\eqn\gff{0\to \Gamma(D_4)\to \Gamma(E_6)\to \Z_3\to 0.}
The other reads 
\eqn\jff{0\to \Gamma(E_6)\to \Gamma(E_7)\to \Z_2\to 0.}
We have here labeled the finite subgroups of $SU(2)$ by the associated
Lie groups.  

These exact sequences determine actions of $\Gamma(E_6)$ and $\Gamma(E_7)$
on $\C^2\times \S^1$ with standard action on $\C^2$ and non-trivial action
on $\S^1$.  (There are no analogous constructions for $E_8$, as $\Gamma(E_8)$
is simple.)
With these actions, $(\C^2\times \S^1)/\Gamma(E_6)$ has a circle
of $D_4$ singularities with $\Z_3$ monodromy.  The monodromy acts by an
outer automorphism of $D_4$ that breaks $D_4$ to $G_2$ (or a 
subgroup that can be obtained by Higgsing).  So in the small radius
limit, we get a six-dimensional theory with $G_2$ gauge group.

On the other hand, $(\C^2\times \S^1)/\Gamma(E_7)$ has a circle of
$E_6$ singularities with a $\Z_2$ monodromy that breaks $E_6$ down
to $F_4$ (or a subgroup that can be obtained by Higgsing).
So in the small radius limit, 
we get a six-dimensional theory with $F_4$ gauge group.

To summarize what we have learned in this section, from
$M$-theory on $(\C^2\times \S^1)/\Gamma$, one can obtain in a small
radius limit, with a suitable action of $\Gamma$ and suitable Wilson
lines, six-dimensional theories with any desired simple gauge group
at low energies.

\newsec{Realization By Matrix Theory}

In this section, we return to the $(p,q)$ fivebrane theory and
describe a matrix model realization.

First we recall from \ref\matu{M. Douglas, ``Enhanced Gauge Symmetry
In M(atrix) Theory,'' hep-th/9612126.} 
\nref\a{M. Douglas and G. Moore, ``$D$-Branes, Quivers, and ALE
Instantons,'' hep-th/9603167.}
\nref\b{C. V. Johnson and R. C. Myers, ``Aspects Of Type IIB
Theory On ALE Spaces,'' Nucl. Phys. {\bf B484} (1997), hep-th/9610104.}
(following \refs{\a,\b})
the description of matrix
theory on $\R^7\times\C^2/\Z_q$.  In matrix theory, one works in light
cone gauge, and so in this section we will write only the transverse
part of the spacetime, which in the present example is $\R^5\times \C^2/\Z_q$.
To describe matrix theory with this transverse spacetime, one needs
a $\Z_q$-invariant collection of zerobranes on $\R^5\times \C^2$.
An important special case is that in which $\Z_q$ acts by cyclic permutations
on $q$-plets of zerobranes.  (Other cases correspond to a matrix description
of states carrying nonzero charges under the maximal torus of the
$SU(q)$ gauge symmetry of $M$-theory on $\R^7\times \C^2/\Z_q$.) 
In this case, the total number of zerobranes is $Nq$ for some $N\geq 0$;
one ultimately wishes to take $N\to \infty$.  

\def\NN{{\bf N}}
\def\oone{\bf 1}
The matrix description of this system is based on a quantum mechanical
model with eight supercharges (which could come, for example, by dimensional
reduction from ${\cal N}=2$ in four dimensions), gauge group
$U(N)^q$, and hypermultiplets
which if the $U(N)$ factors in the gauge group are correctly ordered
transform as $(\NN,\bar \NN,\oone,\dots,\oone)
\oplus (\oone,\NN,\bar \NN,\oone,\dots,\oone)\oplus \dots$ (a sum of
$q$ terms obtained by cyclic permutations).  We will recall below
how this structure comes about.

Classically, this
theory has several branches of vacua, parametrized by the expectation
values of scalar fields $H$ in hypermultiplets and scalar fields
$\Phi$ in the $U(N)^q$ vector multiplets.  In particular, there is
a branch on which the $H$'s have generic expectation values (compatible
with vanishing of the $D$ terms). 
On this branch, which we call the Higgs branch,
the gauge group is broken to $U(1)^N$.
On the Higgs branch, the expectation values of the $H$'s 
parametrize the positions of $N$ zerobranes on $\C^2/\Z_q$,
and the expectation values of the $\Phi$'s parametrize the positions
of the $N$ zerobranes on $\R^5$.  (The unbroken $U(1)^N$ gauge group has
a simple intuitive interpretation: there is a $U(1)$ factor for each
zerobrane.) Altogether, then, this branch is related
to the motion of zerobranes on the transverse space $\R^5\times \C^2/\Z_q$.
There is also, classically, a Coulomb branch on which the $H$'s are zero
and the expectation values of the $\Phi$'s break the gauge group to
a maximal torus $U(1)^{Nq}$; the eigenvalues of the $\Phi$'s can be
 parametrized by the positions of
$Nq $ ``particles'' on $\R^5$.  Quantum mechanically, wave functions
from one branch leak onto another (at least for any fixed $N$), and there
is no precise separation between them.

To consider instead $M$-theory on $\R^6\times \C^2/\Z_q\times \S^1$,
that is on a transverse space $\R^4\times \C^2/\Z_q\times \S^1$,
one must in the usual fashion \ref\wati{W. Taylor, ``$D$-Brane Field
Theory On Compact Spaces,'' hep-th/9611042.} $T$-dualize
in one of the coordinates.  Then one gets a $1+1$-dimensional
$(4,4)$ supersymmetric gauge theory with gauge group $U(N)^q$.
In addition to the vector multiplets of $U(N)^q$, there are hypermultiplets
transforming still as $(\NN,\bar \NN,\oone,\dots,\oone)
\oplus (\oone,\NN,\bar \NN,\oone,\dots,\oone)\oplus \dots$.  The 
$1+1$-dimensional theory is formulated on $\R\times\tilde \S^1$, where
$\tilde \S^1$ is ``dual'' to the $\S^1$ in the original $M$-theory
description.  For reasons we have recalled in section 2, a decoupled
six-dimensional theory is expected to appear in the limit in which
the radius of $\S^1$ goes to zero, and that of $\tilde \S^1$ goes therefore
to infinity.

Like the $0+1$-dimensional system whose behavior was summarized two
paragraphs ago, this system has several branches of classical vacua.  There
is a Higgs branch, in which the $H$'s have generic expectation values,
and a Coulomb branch, in which the $H$'s vanish.  The former describes
motion of $N$ strings on $\R^4\times \C^2/\Z_q$  and the latter describes
the motion of $Nq$  ``strings'' on a transverse $\R^4$. (These statements
are closely related to the corresponding $0+1$-dimensional statements
summarized above; in going from $0+1$ to $1+1$ dimensions, zerobranes
become onebranes or strings and the $\Phi$'s are reduced from
describing a transverse $\R^5$ to a transverse $\R^4$.)  In the limit
that the radius of $\tilde \S^1$ goes to infinity, the Higgs and Coulomb
branches flow to separate conformal field theories, for reasons explained
in \witten.
Evidently, in this limit, if matrix theory is correct, the Higgs branch
describes the free Type IIA theory on transverse $\R^4\times \C^2/\Z_q$,
and the Coulomb branch describes the decoupled six-dimensional 
$SU(q)$
``gauge'' theory with transverse $\R^4$.

\nref\various{O. Aharony, M. Berkooz, S. Kachru, N. Seiberg, E. Silverstein,
  ``Matrix Description of Interacting Theories in Six Dimensions,''
    hep-th/9707079.} 
\nref\nwitten{E. Witten, ``On The Conformal Field Theory Of The Higgs
Branch,''hep-th/9707093.}
The last statement, which has also been obtained in \othersethi,
 is, roughly, ``mirror'' to a corresponding statement for
a matrix theory realization of a six-dimensional theory related to Type IIA
parallel fivebranes.  In that case \refs{\various,\nwitten},
the conventional physics comes from a Coulomb branch and the decoupled
six-dimensional physics comes from a Higgs branch.

\subsec{$(p,q)$ Case}

Now we come to our real interest, which is the $(p,q)$ generalization
explored in sections 2-5.
For this, in view of the discussion in section 3, we must consider $M$-theory
on a transverse $\R^4\times (\C^2\times\S^1)/\Z_q$, where the $\Z_q$ action
is as in  equation \nobb:
\eqn\onobb{\eqalign{z_1 & \to e^{2\pi i/q}z_1\cr
                   z_2 & \to e^{-2\pi i/q}z_2\cr
                   t & \to t-2\pi R{p\over q}\cr}}
                   
I claim that this theory has a matrix realization in terms of
a $(4,4)$ supersymmetric gauge theory on $\R\times \tilde \S^1$
which is a ``twisted'' version of the one just described.
The gauge group is still $U(N)^q$, and (with a suitable cyclic
ordering of the
$U(N)$ factors) the hypermultiplets still transform as 
$(\NN,\bar \NN,\oone,\dots,\oone)
\oplus (\oone,\NN,\bar \NN,\oone,\dots,\oone)\oplus \dots$.
However, as one goes around the circle $\tilde \S^1$, the $q$ factors
in the gauge group $U(N)^q$ are permuted by an outer automorphism that
preserves the cyclic ordering of the factors but moves each factor
$p$ steps to the ``left.''  Thus, locally on $\tilde \S^1$, the theory
is actually independent of $p$.

Just as in the case $p=0$, the model has a Higgs branch and a Coulomb
branch at the classical level.  In the limit that the radius of $\S^1$
goes to zero (and that of $\tilde \S^1$ to infinity), the Higgs branch
and the Coulomb branch become  separated quantum mechanically.  The 
six-dimensional model that we explored in sections 2-5 arises from the Coulomb
branch.

To analyze the problem, it will be helpful to view $X_{p,q}=(\C^2\times\S^1)/
\Z_q$ as $(\C^2\times\R)/\Gamma$, where $\Gamma$ is the discrete group
generated by
\eqn\tonobb{\eqalign{
\alpha:&z_1\to e^{2\pi i/q}z_1,~z_2\to e^{-2\pi i/q}z_2,~t\to t
-{2\pi R\over q}\cr \beta:&z_1\to z_1, ~z_2\to z_2, ~t\to t+2\pi R.\cr}}
Note that $\alpha$ and $\beta $ obey the one relation
\eqn\yonobb{\alpha^q=\beta^{-p}.}

To study $M$-theory on transverse $\R^4\times X_{p,q}$ using
this description of $X_{p,q}$, we must study $\Gamma$-invariant configurations 
of zerobranes on $\R^4\times \C^2\times \R$.

Consider, in general, the action of a discrete group $\Gamma$ on a manifold
$P$ and the problem of describing $\Gamma$-invariant zerobrane configurations
on $P$.  $\Gamma$-invariance means that the zerobrane configuration is
a sum of $\Gamma$ orbits.  If $\Gamma$ acts freely on $P$, each $\Gamma$
orbit is a copy of $\Gamma$ itself.  In this case, the zerobrane positions
are labeled by an element $\gamma\in \Gamma$
\foot{To be pedantic, a free $\Gamma$ orbit has no canonical point
on it, unlike the group $\Gamma$, which has a canonical point, namely
the identity element of the group.  Thus the free orbit should be regarded
as a ``principal homogeneous space'' for $\Gamma$ action, which is
a copy of $\Gamma$ on which $\Gamma$ acts from (say) the left, but
which is not endowed with a group structure.}
 and by possible additional
labels.  If $\Gamma$ does not act freely on $P$, there are also smaller orbits,
supported at fixed points of elements of $\Gamma$.

In our problem, $\Gamma $ acts freely if and only $p$ and $q$ are relatively
prime.  Otherwise, there are fixed points of a $\Z_r$ subgroup of $\Gamma$
($r$ being as before the greatest common divisor of $p$ and $q$)
at the origin in $\C^2$.  Inclusion of the non-free $\Gamma$ orbits
will give, as in \matu, a matrix description of states carrying non-zero
charges under the maximal torus of the unbroken $SU(r)$ gauge group.
For simplicity, we will here consider only the free $\Gamma$ orbits.

Just as in the case of $M$-theory on 
a circle, it is helpful to make $T$-duality with respect to the group $\Gamma$.
This is done by introducing the unitary
representations of the group $\Gamma$,
a procedure that is particularly useful when $\Gamma$ is abelian, as in the
present case.  A unitary representation is determined by setting
$\beta=e^{i\theta}$ for some real theta, and setting $\alpha$ to be
a complex number (which we also call $\alpha$) that obeys
\eqn\ucc{\alpha^q=e^{-ip\theta}.}
Here $\theta$ is a coordinate on a dual circle $\tilde S^1$ on which
the matrix string theory will be formulated.  Since \ucc\ has $q$
solutions for each value of $\theta$, this family of $\Gamma$ representations
is a $q$-fold cover $C$ of $\tilde \S^1$.
$C$ looks locally like $q$ copies of $\tilde \S^1$, but under $\theta\to
\theta+2\pi$, the branches undergo a cyclic permutation by $p$ steps.
We will find (as one would expect from \matu) that the branches correspond
to the factors of $U(N)$ in the gauge group $U(N)^q$; the behavior under
$\theta\to\theta+2\pi$ will lead to an outer automorphism of the gauge
group, cyclically permuting the factors, in going around $\tilde \S^1$.

Let us consider the case that the zerobranes consist of $N$ free $\Gamma$
orbits.  A component of zerobrane position can thus be labeled by
$x_{i\gamma}$, with $\gamma\in \Gamma$ and $i=1,\dots,N$.  For clarity
in the following, we will suppress the $i$ index, except in stating
final results.  Inclusion of the $i$ index has the effect of making everything
$N\times N$ matrix-valued.

Let $x$ be a component of zerobrane position on which $\Gamma$ acts
by $\alpha x=\zeta x$ and $\beta x = x$.  Thus, in terms of the transverse
space $\R^4\times (\C^2\times \R)$, $x$ could be a coordinate on $\R^4$,
in which case $\zeta=1$, or $x$ could be a coordinate $z_1$ or $z_2$
of $\C^2$, in which case $\zeta=e^{\pm 2\pi i/q}$.  

With the $i$ index suppressed, $x$ is a matrix $x_{\gamma,\gamma'}$ in
a basis given by elements of $\Gamma$.  If we write $\gamma=\alpha^a\beta^s$
where we can take $a=0,1,\dots,q-1$, then $x$ is a matrix
$x_{a,s;b,t}$.  To increase the legibility of the formulas,
we will write this as $x(a,s;b,t)$.
The transformation law of $x$ under $\Gamma$ means that
\eqn\unny{\eqalign{x(a+1,s;b+1,t)&=\zeta x(a,s;b,t) \cr
              x(a,s+1;b,t+1)&=x(a,s;b,t). \cr}}
              
Now we make a Fourier
transform from the basis $b,t$ to a basis of functions of $\alpha$
and $\theta$:
\eqn\ununy{x(\alpha',\theta';\alpha,\theta)={1\over 2\pi q}\sum_{b',t';b,t}
x(b',t';b,t)(\alpha')^{-b'}e^{-it'\theta'} \alpha^be^{it\theta}.}
Using \unny, this becomes
\eqn\hunny{x(\alpha',\theta';\alpha,\theta)= {1\over 2\pi q}
\sum_{b',t';b,t} x(b'-b,t'-t;0,0)(\zeta \alpha(\alpha')^{-1})^{b}
(\alpha')^{b-b'}e^{it(\theta-\theta')}e^{i(t-t')\theta'}.}
The sum over $b$ and $t$ for fixed $b-b'$, $t-t'$ now gives delta
functions setting
\eqn\uvvu{\eqalign{ \theta' & = \theta   \cr
                     \alpha' & =\zeta \alpha.\cr}}
The first equation says that $x$ acts locally in $\theta$.  
Reintroducing the so far suppressed index $i=1,\dots, N$, $x$ acts  at
given $\theta$ on a space of dimension $Nq$; a basis of this
space is labeled by $i$ and by
the choice of $\alpha$ with $\alpha^q=e^{-ip\theta}$.  
However, the second equation in
\uvvu\ says that for $\zeta=1$, $x$ acts diagonally on the
$\alpha$ index.  Thus, the $x$'s that have $\zeta=1$ split up at fixed
$\theta$ as the
sum of $q$ different $N\times N$ blocks, one for each value of $\alpha$.       
                 
These $x$'s transform in the adjoint representation of $U(N)^q$, and
are in fact the scalar fields in the vector multiplet.  

On the other hand, if $\zeta=e^{\pm 2\pi i/q}$, then  the second equation in
\uvvu\ says that multiplication by $x$  makes a cyclic permutation on the set
of possible values of $\alpha$ by $\pm 1$ step.  As a result, these components
of $x$, which are the coordinates $z_1, z_2$ of $\C^2$, transform under
the gauge group
as $(\NN,\bar \NN,\oone,\dots,\oone)\oplus \dots$
in the case of $z_1$, or $(\bar \NN,\NN,\oone,\dots,\oone)\oplus \dots$
for $z_2$.  (In each case, $\dots$ refers to terms obtained via
cyclic permuations.)  These fields are the bosonic part of
$1+1$-dimensional hypermultiplets transforming as $(\NN,\bar \NN,\oone,
\dots,\oone)\oplus{\rm cyclic~permutations}$.

It remains to make a similar analysis for the remaining coordinate
$t$, which according to \onobb\ transforms inhomogeneously under $\Gamma$.
In this case, we have 
\eqn\illo{\eqalign{t(\alpha',\theta';\alpha,\theta)&
={1\over 2\pi q}\sum_{b',u';b,u}
t(b',u';b,u)(\alpha')^{-b'}e^{-iu'\theta'} \alpha^be^{iu\theta}\cr &
={1\over 2\pi q} \sum_{b',u';b,u}\left(t(b'-b,u'-u;0,0)+2\pi R\left(
u-{p\over q}b\right)
\right)\cr &~~~~~~~
(\alpha')^{-(b'-b)}e^{-i(u'-u)\theta'} (\alpha(\alpha')^{-1})^b
e^{i(u-u')\theta}.\cr}}
The term on the right which is homogeneous in $t$ gives, after 
summing over $b$ and $u$ for fixed $b'-b$ and $u'-u$, a contribution
proportional to $\delta(\theta'-\theta)\delta_{\alpha',\alpha}$.
This contribution is local in $\theta$ and diagonal in $\alpha$ and
corresponds, after remembering to include the additional
label $i=1,\dots, N$, to a function $A(\theta)$ with values in the
adjoint representation of $U(N)^q$.  This will
be interpreted as the gauge field on $\tilde\S^1$.

The $2\pi Ru$ term on the right
hand side gives after the sum over $u$ a term $-i\delta'(\theta-\theta')$,
which is the matrix element of $-id/d\theta$.  Adding this to the   term
analyzed in the last paragraph would give
a covariant derivative $-iD/D\theta=-d/d\theta+A$.
 At first sight, the
$2\pi Rb(-p/q)$ term on the right of \illo\ is obscure.  However,
we must remember that because of the relation $\alpha^q=e^{-ip\theta}$,
we cannot vary $\theta$ keeping $\alpha$ fixed.  If we change variables
from  $\theta$ and $\alpha$ to $\theta$ and 
$\tilde\alpha=\alpha e^{i(p/q)\theta}$ (which can remain fixed as $\theta$
varies) then in the $e^{iu(\theta-\theta')}$ factor in \illo, $u$ is
replaced by $u-(p/q)b$.  So the inhomogeneous term on the right hand
side of \illo\ is the Fourier transform of $-id/d\theta|_{\tilde\alpha}$,
the  $\theta$ derivative at fixed $\tilde\alpha$.  The net effect
is that $t$ is interpreted as a covariant derivative at fixed
$\tilde \alpha$, 
\eqn\jokery{t\to -i \left.{D\over D\theta}\right|_{\tilde\alpha}.}

Varying $\theta$ at fixed $\tilde\alpha$ means that under $\theta\to\theta+
2\pi$, $\alpha$ is transformed to $\alpha e^{-2\pi i(p/q)}$, as we have
claimed.
This transformation of $\alpha$ in going around $\tilde \S^1$ is the
one essentially new point in our analysis, compared to the familiar
case of $p=0$.  Since the $U(N)$ factors in the gauge group are associated
with the possible values of $\alpha$, this shift
 means that in going around $\tilde \S^1$, the
gauge group $U(N)^q$ is transformed by an outer automorphism that moves
each factor $p$ steps to the left.

\bigskip
\noindent{\it Acknowledgements}
I would like to thank S. Shenker and H. Verlinde for raising questions that
stimulated this study,
D. Morrison for comments and for pointing out reference \reid, H. Ooguri
and S. Sethi for discussions, A. Karch and B. Kol for pointing out an
error in an earlier version of section 5,
and the Aspen Center for Physics for its hospitality.
\listrefs

\end